\documentclass[preprint,12pt]{elsarticle}

\usepackage{amssymb}

\usepackage{hyperref}

\usepackage{bm}
\usepackage{amsmath}
\usepackage{slashed}
\usepackage{xcolor}
\usepackage{graphicx}
\usepackage{caption}
\usepackage{subcaption}
\usepackage[normalem]{ulem}
\usepackage[utf8]{inputenc}
\usepackage[T1]{fontenc} 

\newcommand{\MS}{{\overline{\rm MS}}}

\newcommand{\beq}{\begin{align}}
\newcommand{\eeq}{\end{align}}

\newcommand{\nn}{\nonumber}

\DeclareRobustCommand{\eq}[1]{Eq.~(\ref{eq:#1})}
\DeclareRobustCommand{\eqs}[2]{Eqs.~(\ref{eq:#1}) and (\ref{eq:#2})}

\DeclareRobustCommand{\fig}[1]{Fig.~\ref{fig:#1}}

\journal{Nuclear Physics B}

\begin{document}

\begin{frontmatter}



\title{A Hybrid Renormalization Scheme for Quasi Light-Front Correlations in Large-Momentum Effective Theory}


\author{Xiangdong Ji}
\address{Center for Nuclear Femtography, SURA, 1201 New York Ave. NW, Washington, DC 20005, USA}
\address{Maryland Center for Fundamental Physics,
Department of Physics, University of Maryland,
College Park, Maryland 20742, USA}

\author{Yizhuang Liu}
\address{Tsung-Dao Lee Institute, Shanghai Jiao Tong University, Shanghai, 200240, China}

\author{Andreas Sch\"{a}fer}
\address{Institut f\"{u}r Theoretische Physik, Universit\"{a}t Regensburg, D-93040 Regensburg, Germany}

\author{Wei Wang}
\address{INPAC, Shanghai Key Laboratory for Particle Physics and Cosmology, MOE Key Lab for Particle Physics, Astrophysics and Cosmology, School of Physics and Astronomy, Shanghai Jiao Tong University, Shanghai, 200240, China}

\author{Yi-Bo Yang}
\address{CAS Key Laboratory of Theoretical Physics, Institute of Theoretical Physics, Chinese Academy of Sciences, Beijing 100190, China}
\address{School of Fundamental Physics and Mathematical Sciences, Hangzhou Institute for Advanced Study, UCAS, Hangzhou 310024, China}
\address{International Centre for Theoretical Physics Asia-Pacific, Beijing/Hangzhou, China}

\author{Jian-Hui Zhang}
\address{Center of Advanced Quantum Studies, Department of Physics, Beijing Normal University, Beijing 100875, China}

\author{Yong Zhao}
\address{Physics Department, Brookhaven National Laboratory Bldg. 510A, Upton, NY 11973, USA}

\author{}

\address{}

\begin{abstract}

In large-momentum effective theory (LaMET), calculating parton physics starts from calculating coordinate-space-$z$ correlation functions
$\tilde h(z, a,P^z)$ in a hadron of momentum $P^z$ in lattice QCD. Such correlation functions involve
both linear and logarithmic divergences in lattice spacing $a$, and thus need to be properly renormalized.
We introduce a hybrid renormalization procedure to match these lattice
correlations to those in the continuum $\overline{\rm MS}$ scheme, without introducing extra
non-perturbative effects at large $z$. We analyze the effect of ${\cal O}(\Lambda_{\rm QCD})$ ambiguity in the Wilson line self-energy subtraction involved in this hybrid scheme. To obtain the momentum-space
distributions, we recommend to extrapolate the lattice data to
the asymptotic $z$-region using the generic properties of the
coordinate space correlations at moderate and large $P^z$, respectively.

\end{abstract}




\end{frontmatter}


\section{Introduction}
\label{sec:intro}

Parton physics is important both for understanding the dynamics of
high-energy collisions of hadrons and for studying their
internal structure~\cite{Ellis:1991qj,Thomas:2001kw}. The most familiar examples
are quark and gluon parton distribution functions (PDFs)
which, on one hand, provide the beam information for
high-energy productions at colliders~\cite{Gao:2017yyd}, and on the other hand,
describe the bound-state physics of the colliding hadrons.

Despite its importance, calculating parton physics from first principles of
quantum chromodynamics (QCD) has been a challenge. Recently,
an effective field theory (EFT) approach--large momentum effective theory (LaMET)--has been proposed to extract parton physics from physical properties
of hadrons moving at large momentum~\cite{Ji:2013dva,Ji:2014gla}, where the latter
can be calculated from systematic approximations to Euclidean QCD such
as lattice field theory. Since its proposal, LaMET has been widely used in calculating quark isovector distribution functions~\cite{Lin:2014zya,Alexandrou:2015rja,Chen:2016utp,Alexandrou:2016jqi,Alexandrou:2018pbm,Chen:2018xof,Lin:2018qky,Liu:2018uuj,Alexandrou:2018eet,Liu:2018hxv,Chen:2018fwa,Izubuchi:2019lyk,Shugert:2020tgq,Chai:2020nxw,Lin:2020ssv,Fan:2020nzz}, distribution amplitudes (DAs)~\cite{Zhang:2017bzy,Chen:2017gck,Zhang:2020gaj},
generalized parton distributions~\cite{Chen:2019lcm,Alexandrou:2019dax}, and recently transverse-momentum-dependent
distributions~\cite{Shanahan:2019zcq,Shanahan:2020zxr,Zhang:2020dbb}, and even higher-twist distributions~\cite{Bhattacharya:2020cen}. Some recent reviews on
LaMET can be found in Refs.~\cite{Ji:2020ect,Cichy:2018mum}.

The key idea of LaMET is that partons in the infinite momentum frame (IMF)
can be approximated by physical properties of a hadron at large but
finite momentum. Due to the existence of ultraviolet (UV) divergences, this approximation
is not completely straightforward. It requires using the standard EFT technology of matching and running. Detailed investigations have shown that the standard DGLAP evolution~\cite{Dokshitzer:1977sg,Gribov:1972ri,Altarelli:1977zs} has its origin in the momentum evolution of physical properties of the hadron~\cite{Ji:2020ect}.

In LaMET applications, one begins with lattice calculations of spatial correlation functions. Since lattice
breaks the continuum symmetry, power divergences
appear in bare correlation functions. They must be subtracted
when matched to those in a continuum scheme such as dimensional regularization
and (modified)-minimal subtraction, ${\overline {\rm MS}}$. In the past, the main
approaches suggested in practical applications include the regularization-independent momentum subtraction method
or RI/MOM~\cite{Constantinou:2017sej,Stewart:2017tvs,Alexandrou:2017huk,Chen:2017mzz} and the
ratio method~\cite{Radyushkin:2017cyf,Orginos:2017kos,Braun:2018brg,Li:2020xml}. The latter relies on
the validity of Euclidean operator product expansion (OPE) and can only be applied to
correlations at short distances, and therefore cannot be used directly for
LaMET applications.
In contrast, the RI/MOM method appears to be applicable to large-$z$ ($z$ is the gauge-link length)
distance at first glance. However, a detailed examination shows that this method introduces potential
non-perturbative effects, for instance, through infrared (IR) logarithms
such as $\ln(z^2\mu^2)$ ($\mu$ is a renormalization scale) in the scheme matching. Since UV divergences are supposed to be
perturbative in asymptotically-free theories such as QCD, it shall be possible to find a renormalization procedure which does not introduce  non-perturbative effects. It might be possible that RI/MOM does not introduce a large non-perturbative effect in the present precision
of lattice-QCD calculations. However, a systematic effective-theory
calculation with high precision cannot avoid addressing this issue.

To achieve this, we propose in this paper a hybrid renormalization procedure for lattice correlations in LaMET applications. At short distances
where OPE is valid,
the standard RI/MOM or ratio method is recommended.
At large distances, we suggest to use the auxiliary field formalism~\cite{Samuel:1978iy,Gervais:1979fv,Arefeva:1980zd,Dorn:1986dt} which has been advocated in LaMET
applications by a number of authors~\cite{Ji:2017oey,Green:2017xeu,Green:2020xco}.
In this formalism, the Wilson line is replaced by two-point functions of the auxiliary field. The linear divergence in
lattice correlation functions is then linked to the mass
renormalization of the auxiliary field, whereas the
logarithmic divergence appears in the renormalization of
the ``heavy''-light ``currents'' at the end of the Wilson line.
Both divergences can be separately renormalized in a manner which is consistent
with the $\overline{\rm MS}$ scheme~\cite{Green:2017xeu,
Green:2020xco}.  Although
the mass subtraction of the Wilson
line has been suggested before~\cite{Chen:2016fxx,Ishikawa:2017faj},
it has not been put into wide practical use because, to our knowledge,
a reliable approach to calculate the non-perturbative mass has not been well-established
in the literature. Here we suggest several ways to do so which shall be investigated through systematic
lattice simulations in the future.

In addition, we also address several other issues that are important in
extracting parton physics using LaMET, e.g., how to match appropriately to the continuum scheme near $z\sim 0$, and how to utilize the asymptotic behavior of relevant correlation functions at large light-front (LF) distance to remove the unphysical oscillations in the momentum distribution that arise from truncated Fourier transform.

\section{Partons as quanta in infinite-momentum states and
large-momentum expansion}
\label{sec:imf}

Let us begin with a brief overview of the parton formalism. In the textbooks, PDFs are usually defined in terms of
LF correlations in QCD~\cite{Sterman:1994ce,Collins:2011zzd}.
The LF is defined by $t-z=$~constant, if a massless particle
is travelling along the $z$-direction, with variations of other
coordinates, $t+z$ and transverse-space dimensions, defining
a three-dimensional front surface.
Introduce two independent LF four-vectors with dimension-one
parameter $\Lambda$,
\begin{eqnarray}
   p^\mu &=& (\Lambda, 0, 0, \Lambda )\ , \nonumber \\
   n^\mu &=& \left(1/(2\Lambda), 0, 0, -1/(2\Lambda)\right)\ ,
\end{eqnarray}
then $p^2=n^2=0$, and $p\cdot n=1$. Different LFs
are defined by different coordinate distance $\lambda$
along the $n$-direction.

Consider now the quark PDFs in a state
$|P\rangle$ with mass $M$ and four-momentum $P^\mu= (P^0,0,0,P^z)
= p^\mu + (M^2/2) n^\mu$, which can be used to solve for $\Lambda =(P^0+P^z)/2$.
Using $\psi$ to denote a full-QCD quark field,
the LF correlation function in coordinate space is,
\begin{equation}
     h(\lambda)= \frac{1}{2P^+}\langle P|\overline \psi(\lambda n)\Gamma W(\lambda n, 0)\psi(0)|P\rangle \ ,
\label{Eq:correlationm}
\end{equation}
where $\Gamma$ is a Dirac matrix, $W$ is a straight Wilson-line gauge link,
and $\lambda$ is the LF distance. All other coordinates have been taken
to be zero. Due to the invariance of the LF under Lorentz boosts along the $z$-direction, the
above correlation function is independent of the residual momentum $P^z$. Quite often, $P^z$ is taken to be zero.

The quark PDF is just the Fourier transform of the above LF correlation~\cite{Collins:2011zzd},
\begin{equation}
     f(x) =\int^\infty_{-\infty} \frac{d\lambda}{2\pi} e^{-ix\lambda}h(\lambda)  \ .
\end{equation}
In this way, partons can be studied without using the EFT machinery although they are effective degrees of freedom (dof's)
to describe the LF collinear modes. The reason is that, the parton dof's are automatically projected out through the LF correlators applied to the full QCD state $|P\rangle$.
On the other hand, these parton dof's can also be explicitly separated in the QCD Lagrangian, as is done in soft-collinear effective theory (SCET) where they are represented by LF collinear 
fields~\cite{Bauer:2000yr,Bauer:2001ct,Bauer:2001yt}.

In the traditional parton formalism, the correlations are time-dependent, or in other words, the operators are in the Heisenberg picture. As such, we say that the formalism
is Minkowskian and thus difficult for Monte Carlo simulations due to
the famous ``sign'' problem. If one chooses
$\xi^- = (t+z)/\sqrt{2}$ as the ``new time'' coordinate,
and integrates it out, one obtains a Hamiltonian formalism
for partons, which has been called LF quantization (LFQ) in the
literature~\cite{Dirac:1949cp}. LFQ
is also a very difficult formalism to work with, despite the fact that
much progress has been made~\cite{Brodsky:1997de}.

An alternative parton formalism can be obtained
by adapting Feynman's original idea about partons
to the context of a field theory~\cite{Ji:2014gla,Ji:2020ect}.
Feynman considered~\cite{Feynman:1973xc} the momentum distribution of a composite system,
$f(k^z, P^z)$, where $P^z$ is the center-of-mass momentum and $k^z$ the longitudinal momentum carried by the parton whose
transverse momentum $\vec{k}_\perp$ has been integrated over.
The $P^z$-dependence of the momentum distribution is clearly
a relativistic effect: According to Poincar\'e symmetry,
the Hamiltonian of a system depends on the frame, and changes under Lorentz boosts
according to,
\begin{equation}
  [H, K^i] = i P^i\ ,
\end{equation}
where $K^i$ $(i = 1,2,3)$ are the boost operators. Therefore,
the wave functions are frame-dependent, leading to frame-dependent
momentum distributions. Because $K^i$ depends on interactions, the
frame-dependence is a dynamical problem, and generally requires
non-perturbative solutions.

An important feature of the momentum distribution of a system is that it is a static or time-independent quantity. In QCD,
it is related to the following spatial correlation,
\begin{equation}
    \tilde h(z,P^z)= \frac{1}{2N}\langle P|\overline \psi(z)\Gamma W(z, 0)\psi(0)|P\rangle \ ,
\label{eq:esme}
\end{equation}
where $W(z,0)$ is a spacelike, straight-line gauge link, and $N$ is normalization factor depending on the Dirac matrix.
Feynman then considered the infinite-momentum limit, assuming that such a limit
exists,
\begin{equation}
  P^z \to \infty, ~~~~~z\to 0, ~~~~~\lambda = zP^z ~{\rm finite}\ ,
\end{equation}
i.e., the relevant correlation function for partons is
\begin{equation}
  \tilde h(\lambda= zP^z) = \langle P^z=\infty|\overline \psi(z)\Gamma W(z, 0)\psi(0)|P^z=\infty\rangle\ .
\label{Eq:correlatione}
\end{equation}
It is clear that in field theories this is a non-trivial limit. In fact,
it can be shown that such a limit only exists in asymptotically-free theories, where
the high-momentum modes are perturbative~\cite{Ji:2020ect}.

If one ignores the subtlety of the limit,
the correlation in Eq.~(\ref{Eq:correlatione})
is related to that  in Eq.~(\ref{Eq:correlationm})
by an infinite Lorentz transformation~\cite{Ji:2013dva}.
Our ``new'' form of parton formalism works with time-independent
correlators and the IMF wave function.
Since the operator is time-independent, it is the
Schr\"odinger representation of parton physics
if an analogy between time-translation and Lorentz boost is
made~\cite{Ji:2020ect}.

In QCD, however, the correlations $\tilde h(\lambda)$
and $h(\lambda)$ are different. The difference arises
from the presence of the UV cut-off. In the physical
momentum distribution, the cut-off must always be much
larger than the hadron momentum. As a result, the parton momentum
is allowed to be larger than the hadron momentum, or $|x|$ can be
larger than 1, without
violating any laws of physics. On the other hand, the standard PDFs
have support $|x|\le1$, corresponding to a UV cut-off smaller than
the hadron momentum. Thanks
to the asymptotic freedom, these two different UV limits
can be connected to each other by perturbation theory in QCD. This makes it possible to extract LF parton physics defined in Eq. (\ref{Eq:correlationm}) from the Euclidean form in Eq. (\ref{Eq:correlatione}).

Eq. (\ref{Eq:correlatione}) is the starting point
of the LaMET expansion, where we first
compute the quasi-LF correlation functions at a finite,
but large momentum $P^z\gg\Lambda_{\rm QCD}$.
To make the expansion work, in principle one needs ${\tilde h}(z, P^z)$ with $-\infty<z<\infty$, or
${\tilde h}(\tilde \lambda, P^z)$ at all quasi-LF distances $\tilde \lambda=z P^z$.
While in reality, of course, due to
the finite volume, lattice data will always stop at some
large $z(\tilde \lambda)$ which we call $z_{\rm L}(\tilde \lambda_{\rm L})$. We will
deal with issues of finite $\tilde \lambda_{\rm L}$ later.
For the discussion in this section, we assume that $\tilde h(\tilde \lambda, P^z)$
is known in $[-\infty,\infty]$, i.e., in the whole $\tilde \lambda$ range at a large
$P^z$.

With the above quasi-LF correlation, one can make a straightforward
Fourier transformation
\begin{equation}
    {\tilde f}(y,P^z) =\int^\infty_{-\infty} \frac{d\tilde \lambda}{2\pi}\, e^{i\tilde \lambda y}\, \tilde h(z, P^z)\ .
\end{equation}
The physical interpretation of ${\tilde f}(y,P^z)$ hinges on the large momentum
expansion~\cite{Ji:2013dva,Xiong:2013bka,Ma:2014jla,Izubuchi:2018srq}
\begin{align}\label{matching}
    {\tilde f}(y, P^z) &= \int^1_{-1} dx\, C\Big(\frac y x, \frac{xP^z}{\mu}\Big) f(x, \mu) \nonumber \\
    &+ {\cal O}\Big(\frac{\Lambda^2_{\rm QCD}}{y^2(P^z)^2}, \frac{\Lambda^2_{\rm QCD}}{(1-y)^2(P^z)^2}\Big),
\end{align}
where $\mu$ is a factorization scale, $\Lambda_{\rm QCD}$ is the hadronic scale, and
$f(x,\mu)$ is the standard PDF that can be extracted from the above equation.
The large scales $(yP^z)^2$ and $((1-y)P^z)^2$ are associated with the active quark
and the spectator momenta, respectively. According to the standard EFT methodology, 
any large scale that is not forbidden shall be allowed in the expansion, and the linear
dependence is absent in dimensional regularization due to space-time symmetry.    
Therefore, the validity of this expansion relies on the smallness of the expansion
parameters $\Lambda^2_{\rm QCD}/[y^2(P^z)^2]$ and $\Lambda^2_{\rm QCD}/[(1-y)^2(P^z)^2]$. 

The $C$ factor in the above equation can be calculated perturbatively. At leading-order in $\alpha_s$, we can identify $\tilde f(y, P^z)$ with $f(y,\mu)$ (ignoring the power corrections for the moment), thus they have the same asymptotic behavior as $y\to 0$ and $y\to1$. Beyond leading-order, this will be changed by perturbative corrections. To see this, let us take the following simple form of $f(x, \mu)$ as an example
\begin{equation}
x^a (1-x)^b,
\end{equation}
with $a, b$ controlling the asymptotic behavior at $x\to 0$ and $x\to 1$, respectively. The perturbative one-loop corrections lead to the following change in the asymptotic behavior
\begin{align}
\delta\tilde f(y, P^z)\sim \alpha_s y^a \ln y \ \  {\rm as}\ \  y\to 0\,.
\end{align}
When resummed to all orders in perturbation theory, this yields a power law behavior of the form $\tilde f(y, P^z)\sim  y^{a+\gamma}$ with $\gamma$ being associated with the anomalous dimension of the operator defining $\tilde f(y, P^z)$. Similar behavior also occurs as $y\to 1$ for realistic PDFs with $b>0$. This can also be seen from the coordinate space analysis to be presented below.

Therefore, for a given large $P^z$,
there is a range of $y$ where high-order corrections as well as power corrections are small, and this range can be translated into
a valid range $x$ for the PDFs. Thus, one can systematically obtain the PDFs in an interval $[x_{\rm min}, x_{\rm max}]$ ($x_{\rm min}$ will approach 0 and $x_{\rm max}$ approach 1 as $P^z\to \infty$).
In other words, the LaMET
expansion provides a natural way to calculate parton distributions
in an interval of the parton momentum $x$, similar to extracting parton distributions from
experimental data at finite energies.

\section{A hybrid renormalization procedure}
\label{sec:hybrid}

As explained in the previous section,
the LaMET expansion starts from calculating the coordinate-space
correlation functions $\tilde h(z,P^z)$ at large momentum $P^z$ and for the whole range
of distance $-\infty<z<\infty$.
On a discrete
lattice with spacing $a$, the nonlocal quark bilinear operator that defines $\tilde h(z,P^z)$ in \eq{esme} can be multiplicatively renormalized as~\cite{Ji:2017oey,Ishikawa:2017faj,Green:2017xeu}
\begin{align}\label{eq:ren}
&\left[\bar{\psi}(z)\Gamma W(z,0)\psi(0)\right]_{\rm B} \nn\\
&= e^{\delta m|z|}Z(a) \left[\bar{\psi}(z)\Gamma W(z,0)\psi(0)\right]_{\rm R}\,,
\end{align}
up to lattice artifacts~\cite{Constantinou:2017sej,Chen:2017mie}. Here the operator on the l.h.s. is defined in terms of bare fields and couplings, denoted by the subscript ``B'', while the operator on the r.h.s. is renormalized and denoted by the subscript ``R''. Without an explicit
statement, we always assume that the renormalized correlations are eventually defined in the $\MS$ scheme before a Fourier transformation is made to the momentum space.
There are both $z$-independent
logarithmic and $z$-dependent linear divergences.
The former arises from the renormalization
of quark and gluon fields as well as the vertices at the endpoints of the Wilson line, which is included in the factor $Z(a)$, while
the latter comes from the Wilson-line self-energy, which is factored into the exponential $e^{\delta m|z|}$ with $\delta m$ being the ``mass correction'' .

A number of
proposals have been made in the literature~\cite{Ishikawa:2016znu,Chen:2016fxx,Monahan:2016bvm,Radyushkin:2017cyf,Constantinou:2017sej,Green:2017xeu,Stewart:2017tvs,Braun:2018brg,Li:2020xml} to renormalize the above lattice correlation functions $\tilde h(z, a, P^z)$, among which the RI/MOM scheme has frequently
been used~\cite{Constantinou:2017sej,Stewart:2017tvs}. In this approach,
one calculates the matrix elements (amputated Green's function) $Z(z,-p^2,a)$ of the bilocal operators $O(z,a)$ in a deep Euclidean state with momentum squared $-p^2\gg\Lambda_{\rm QCD}^2$ in a
fixed gauge, and then defines $\overline{\rm MS}$ operators
as,
\begin{align}
             O_{\overline{\rm MS}}(z,\mu) \equiv Z_{\overline{\rm MS}}(z,-p^2,\mu)\frac{O(z,a)}{Z(z,-p^2,a)},
\end{align}
where $Z_{\overline{\rm MS}}$ converts the RI/MOM renormalized result to the $\overline{\rm MS}$ scheme. The gauge and $-p^2$ dependences cancel between two $Z$-factors.
The r.h.s. has a proper continuum limit $a\to 0$ without divergences.

However, while the RI/MOM approach is justified
for local operators, it has potential problems when
applied to nonlocal ones.
For instance, when $z$ becomes large,
$Z_{\overline{\rm MS}}(z,-p^2,\mu^2)$ contains IR
logarithms of $z$ and the perturbative calculation
of $z$-dependence is not reliable. Moreover,
although the RI/MOM factor $Z(z,-p^2,a)$ helps to cancel
the lattice UV divergences, the composite operator
at large-$z$ contains non-perturbative physics as well.
Therefore, both $Z$-factors contain non-cancelling
non-perturbative effects which alter the IR properties
of $O(z)$. Thus, the RI/MOM renormalization scheme is not reliable
at large-$z$. Moreover, when gluon distributions
are involved, it requires external off-shell gluon states which
bring in potential mixing with gauge-variant operators and make things much more complicated~\cite{Wang:2019tgg}.

In addition to the renormalization issues at large distances, there are also 
subtleties for renormalization at short distances. While the standard renormalization
of a bilocal operator makes it finite at any non-vanishing
$z$, it becomes divergent in the $z\to 0$ limit. On the other hand, if one performs a resummation of the large 
logarithms at small $z$, the result vanishes at $z=0$. However, the lattice
result at $z=0$ approaches to the matrix element of the vector current. 
Clearly, the two limits, $a\to 0$ and $z\to 0$,
are not interchangeable.

To resolve these issues, we propose in this section a
hybrid scheme to renormalize the correlation functions. The key point of this scheme is that we separate the correlations
at short and long distances and renormalize them separately, and match both procedures
at an intermediate distance $z_{\rm S}$. The
matching point must lie within $[0,z_{\rm LT}]$ where
the leading-twist (LT) approximation for the correlation
operator is valid. Discussions on the value of
$z_{\rm LT}$ can be found in Sec. V.

\subsection{Renormalization at short distance $0\le |z|\le z_{\rm S}$}
\label{sec:short}

To renormalize $\tilde h(z,a, P^z)$ for $0\le |z|\le z_{\rm S}$, particular attention shall be paid to the behavior of the correlation functions in the limit $z\to 0$.

In the continuum $\MS$ sheme, the $z\rightarrow 0$ limit is not smooth and
additional logarithmic UV divergences $\sim \ln z^2$ arise which
when resummed yield zero.  However, this is not the case for the lattice matrix element
$\tilde h(z,a, P^z)$. For finite lattice spacing and non-vanishing $z$,
$\tilde h(z,a, P^z)$ includes UV divergences related to the wave function
renormalization of the bare fields, of the form
$\alpha_s(a)\ln (z^2/a^2)$.
At small $z$, particularly when $z=0$ or $a$, $\tilde h(z,a, P^z)$ has discretization effects
and is related to the
lattice-regulated local matrix element ${\bar\psi}\Gamma \psi$.
In particular, when $\Gamma=\gamma^\mu$,
${\bar\psi}\gamma^\mu \psi$ is conserved and its matrix element is finite
in the $a\to 0$ limit. A function demonstrating this interesting
interplay between lattice regulator and small physical distance
is $\ln [(z^2+a^2)/a^2]$.

The above discrepancy in the small-$z$ regime can be removed through a perturbative conversion between lattice regularization and the continuum $\MS$ scheme, which, however, is known to converge slowly. Instead, a more efficient strategy is to cancel the $\ln z^2$-dependences through lattice renormalization, which corresponds to a scheme ``X'' that is different from $\MS$.
As long as $z$ is in the leading-twist region $|z|\le z_{\rm S}$ where $z_{\rm S}$ is
smaller than $z_{\rm LT}$, the difference between the X-scheme and $\MS$ can be calculated in perturbation theory.

For example, the X-scheme can be implemented by forming the ratio of $\tilde h(z,a, P^z)$ and another
matrix element of the same operator $O(z,a)$,
\begin{align}\label{eq:ratio}
 \frac{\tilde h(z,a,P^z)}{Z_X(z,a)}\,,~~~~~~{\rm for\ }|z|\le z_{S}\,,
\end{align}
where the renormalization factor $Z_X$ corresponds to different choices of the matrix element.
Possible choices for $Z_X$ include
\begin{itemize}
\item{Amputated Green's function of $O(z,a)$ in a single-particle deep Euclidean state, fixed in a particular gauge, e.g., Landau gauge, which defines the RI/MOM-type of schemes~\cite{Constantinou:2017sej,Stewart:2017tvs,Alexandrou:2017huk,Chen:2017mzz}.}
\item{Matrix element of $O(z,a)$ in a hadron state with $P^z=0$, depending on applications~\cite{Radyushkin:2017cyf,Orginos:2017kos}.}
\item{Vacuum ($|\Omega\rangle$) expectation value of $O(z,a)$~\cite{Braun:2018brg,Li:2020xml}.}
\end{itemize}
In the second and third option, the matrix elements are gauge invariant,
and therefore no gauge fixing is needed. For the third option, the quantum numbers of the operator must be the same as those of the vacuum.
As discussed above, $z_{\rm S}$ has to be smaller than $z_{\rm LT}$, which is estimated in Sec. V to be about
$0.25 \sim 0.33$~fm. Of course, the stability of the final result with respect to small variations of $z_{\rm S}$ shall be explicitly verified.

Due to the multiplicative renormalizability of the operator $O(z,a)$, all UV divergences cancel in the ratio in \eq{ratio}, thus allowing us to take the continuum limit,
\begin{align}\label{eq:ratio2}
\lim_{a\to0} \frac{{\tilde h}(z,a,P^z)}{Z_X(z,a)} = \frac{{\tilde h}(z,\epsilon,P^z)}{Z_X(z,\epsilon)} \equiv\tilde h^X(z,P^z)\,,
\end{align}
where the term after the first equal sign refers to a $\MS$
calculation of the same ratio with $\epsilon$ corresponding to dimensional regularization $d=4-2\epsilon$ in the continuum theory. In the limit $|z|\to0$, the $\ln z^2$-dependence is independent of the external state, so it cancels in the ratio, making the latter finite at $z=0$.
Moreover, in the leading-twist region $z\le z_{\rm S} \le z_{\rm LT}$, we can perturabtively match the ratio for any X-scheme to the LF correlation $h(\lambda,\mu)$ through the coordinate-space factorization formula~\cite{Izubuchi:2018srq,Radyushkin:2017lvu}
\begin{align}\label{eq:coordfact}
\tilde h^X(\lambda, P^z) & =\int_{0}^1 d\alpha\ {\cal C}^X\Big(\alpha,\frac{\lambda^2\mu^2}{(P^z)^2}\Big) h(\alpha\lambda,\mu) \nn\\
&\qquad + {\cal O}(z^2\Lambda_{\rm QCD}^2)\ ,
\end{align}
where ${\cal C}^X$ is the matching coefficient, and we have suppressed its dependence on the renormalization scale in the X-scheme such as the RI/MOM. Also, higher-twist contributions have been suppressed in the above equation.

The one-loop matching coefficient for the second option for $Z_X$ has been obtained in Refs.~\cite{Zhang:2018ggy,Radyushkin:2017lvu,Izubuchi:2018srq}, and that for the RI/MOM scheme can be extracted from Ref.~\cite{Constantinou:2017sej,Izubuchi:2018srq}. The two-loop results for both the second and third option have been calculated as a series expansion in Ref.~\cite{Li:2020xml}.

\eq{coordfact} also has an equivalent form in momentum space~\cite{Izubuchi:2018srq}, with the matching coefficient calculated at one-loop~\cite{Stewart:2017tvs,Liu:2018uuj,Liu:2018hxv} and two-loop~\cite{Chen:2020ody} orders.
The two-loop matching coefficient for $Z_X$ defined by $P^z=0$ matrix element can be extracted from Refs.~\cite{Chen:2020ody,Li:2020xml}.

\subsection{Renormalization at large distances $z\ge z_{\rm S}$}
\label{sec:long}

At large distance $z\ge z_{\rm S}$, UV renormalization
needs a careful assessment because both the RI/MOM and the ratio scheme will introduce
undesired non-perturbative effects.
The only renormalization approach that will not introduce such extra non-perturbative physics
is the explicit and separate subtraction of linear divergences (or $\delta m$)
and logarithmic divergences~\cite{Chen:2016fxx,Ji:2017oey,Ishikawa:2017faj,Green:2017xeu},
which in principle can be done using the auxiliary field method~\cite{Green:2017xeu,Green:2020xco}.

To calculate the mass renormalization $\delta m$
of the Wilson line, there exist many
suggestions in the literature. Here we provide probably an incomplete list:

\begin{itemize}
\item{One can fit the hadron matrix element
at large $z$, where the dominant decay is
\begin{equation}\label{eq:mass1}
  \tilde h(z,a, P^z) \sim \exp(-\delta m |z|) \ .
\end{equation}
$\delta m$ can be obtained by fitting the ratio $\ln (\tilde h(z+a,a, P^z)/\tilde h(z,a, P^z))$
to a constant in $z$ at large $z$. Of course, the result has to be independent of $P^z$,
e.g., one can choose $P^z=0$. Alternatively, one can fit $\ln \tilde h(z,a, P^z)$ to the $1/a$
dependence all $z$. This method has yet to be studied using real lattice data.}
\item{One can use the single-quark Green's function
as in the RI/MOM renormalization factor,
\begin{align}
          Z(z,-p^2,a)&=\int d^4x d^4y\ e^{ip\cdot (x-y)}\nn\\
          &\times  \langle \Omega|T\psi(x) O(z,a) \bar{\psi}(y)|\Omega\rangle\,,
\end{align}
which asymptotically goes like $Z(z) \sim \exp(-\delta m |z|)$. This matrix element needs a fixed gauge,
and has been studied in Refs.~\cite{Izubuchi:2019lyk,Huo:2019vdl}.}
\item{On can also use the vacuum matrix element of $O(z,a)$
\begin{equation}
        S(z) = \langle \Omega|O(z,a)|\Omega\rangle
\end{equation}
which is gauge invariant. $S(z)$ again at very large $z$ behaves like
$S(z) \sim \exp(-\delta m |z|)$. This has been considered in Refs.~\cite{Braun:2018brg,Li:2020xml}.
}
\item{Also the gauge-invariant Polyakov loop leads to the static potential between two heavy quarks. There exists a large number of references on this approach, see, e.g.,~\cite{Appelquist:1977tw,Schroder:1999xf,Smirnov:2009fh,Philipsen:2002az,Jahn:2004qr}.}
\item{One can also calculate the vacuum expectation value
of the Wilson line $W(z)$ directly in a fixed gauge,
and again $\langle W(z) \rangle \sim \exp(-\delta m |z|)$ at large distance. This has been considered in Refs.~\cite{Green:2017xeu,Green:2020xco} using the auxiliary field method.}
\end{itemize}
It is worth pointing out that, although all the proposals above work in principle, different practical issues may arise in their lattice realization such that some may work better than the others.

The mass renormalization $\delta m$ is gauge-independent, just like the pole mass
of a quark~\cite{Kronfeld:1998di}. In the above suggestions where no gauge fixing is
needed, this is obviously true. In the cases where a gauge-fixing is needed,
one can demonstrate that the results in any other gauge are the same by
constructing appropriate gauge-invariant operators~\cite{Philipsen:2001ip}.
Despite being gauge-independent, $\delta m$ will depend on the specific action used in Monte Carlo simulations
and on the definition of the matrix elements above.
In the cases of vacuum matrix elements, $\delta m$ may be interpreted
as the non-perturbative pole mass in certain gauges~\cite{Philipsen:2001ip}.

The $\delta m$ calculated from all the matrix elements above will have the following dependence
on the lattice spacing $a$,
\begin{equation}
   \delta m = m_{-1}(a)/a + m_0 \ ,
\end{equation}
where $m_{-1}$(a) is the coefficient of the power divergence, which is independent of the specific matrix element. The $a$-independent term $m_0$
has a more complicated origin. It can arise
from various sources:
\begin{itemize}
\item{Renormalon effect: In principle, $m_{-1}(a)$ can be calculated perturbatively, and is $2\pi\alpha_s(a)/3$ at leading
order, similar to the mass counterterm in the Wilson
formulation of fermions. However, the perturbation series is not convergent. When truncated at order $n\sim 1/\alpha_s(a)$, the
perturbation series has an uncertainty of order $a\Lambda_{\rm QCD}$, which generates a contribution to $m_0$~\cite{Ji:1995tm,Beneke:1998ui,Bauer:2011ws,Bali:2013pla}.
This means that in non-perturbative fitting, $m_{-1}(a)$ is determined only
with an uncertainty of the order of $a\Lambda_{\rm QCD}$. Thus, an additional 
$m_0$ contribution of order $\Lambda_{\rm QCD}$ is expected.}
\item{Pole mass: For certain matrix elements, like vacuum elements of bilocal operators, the $z$
dependence can be viewed as originating from the pole mass of a meson consisting of an infinitely-heavy quark
and a light one. In this case, $\delta m$ is the pole mass apart from the linear divergence.}
\item{Finite $P^z$ effects: The correlation function at finite $P^z$ has a long-range correlation,
$\exp(-\lambda/\xi(P^z))$, where $\xi(P^z)$ is the correlation length. This contribution is included in $m_0$.}
\item{Fitting effect. Since the data is always in finite $z$ and $a$ where the exponential decay
cannot always be separated from an algebraic decay, there are fitting uncertainties
contributing to $m_0$ as well as the separation between $m_0$ and $m_{-1}$.}
\end{itemize}
To summarize, $m_0$ depends on the lattice matrix-element used and the fitting procedure~\cite{Zhang:2017bzy,Chen:2017gck,Izubuchi:2019lyk,Huo:2019vdl}.

\begin{figure}[tbp]
\centering
\includegraphics[width=0.9\columnwidth]{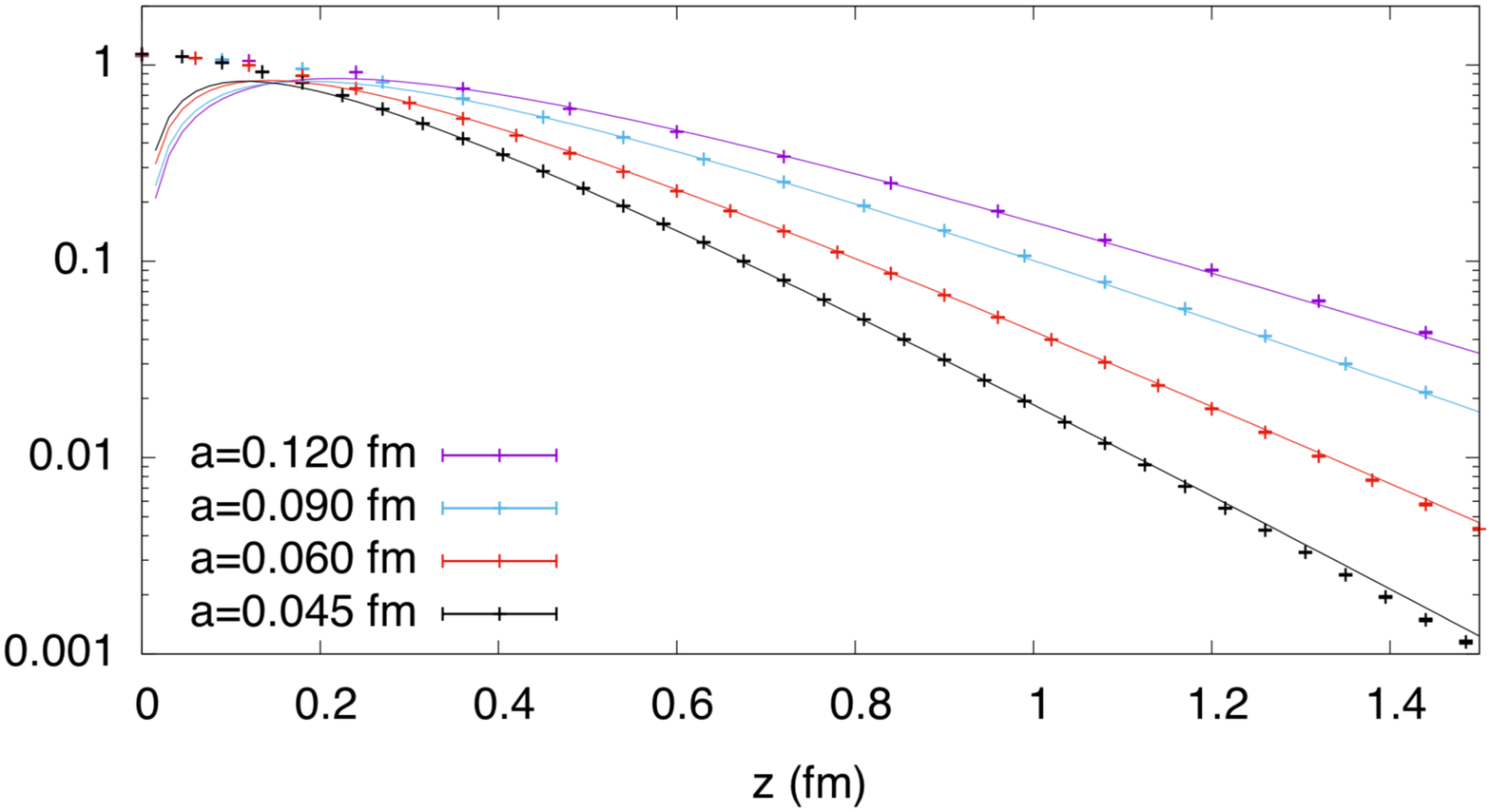}
\vspace*{-2em}
\caption{Fitting of the quark RI/MOM renormalization factor calculated using four ensembles with lattice spacings $a\approx \{0.045, 0.06, 0.09, 0.12\} {\rm fm}$ and $310$ MeV pion mass from MILC collaboration~\cite{Bazavov:2012xda}. }
\label{fig:deltam}
\end{figure}

In Fig.~\ref{fig:deltam} we show, as an example, the values of $m_{-1}$ and $m_0$ determined from the quark RI/MOM renormalization factor calculated at the scale $\mu_R=1.8~ {\rm GeV}$ and $p^z_R=0$, using the four ensembles with $a\approx \{0.045, 0.06, 0.09, 0.12\} {\rm fm}$ and $310$ MeV pion mass from MILC collaboration~\cite{Bazavov:2012xda}. Inspired by the asymptotic behavior at large $z$ to be studied in Sec.~\ref{sec:regge}, we use the following simplified form
\begin{equation}
e^{-(\frac{m_{-1}}{a}+m_0)|z|}\frac{c_1}{|z|^{d_1}}
\end{equation}
to fit the renormalization factors at four different lattice spacings. It is worth pointing out that $m_{-1}$ starts from $O(\alpha_s)$ and we therefore also include the dependence of the coupling on $a$ in the fitting. For $a\approx 0.12$ fm, the fitted results for the coefficients $m_{-1}$ and $m_0$ are
\begin{equation}
m_{-1}=0.234\pm 0.012, \ \ m_0=(350\pm 60)\, {\rm MeV}.
\end{equation}

In principle, one can choose to subtract the power divergent piece only, namely $m_{-1}/a$.
The less-well-determined $m_0$ term can be left in the lattice matrix elements, the
momentum expansion will take care of the rest. Indeed, the difference between subtracting
different $m_0$
is $O(1/P^z)$ effect, as has been demonstrated
perturbatively in~\cite{Xiong:2013bka}.
More precisely, assuming there are two quasi-LF correlations that define the quasi-PDFs and differ from each other by a factor  $e^{-m|z|}$ with $m \sim \Lambda_{\rm QCD}$,
\begin{equation}
\tilde h_1(z,P^z)=\tilde h_2(z,P^z)e^{-m|z|} \ ,
\end{equation}
then after Fourier transforming into momentum space, they are related by
\begin{align}
&\tilde f_1(y,P^z)-\tilde f_2(y,P^z)\nonumber \\
&=\frac{1}{\pi}\int_{-\infty}^{\infty} dy' \frac{\delta}{\delta^2+(y-y')^2} \left[\tilde f_2(y',P^z)-\tilde f_2(y,P^z) \right]\ ,
\end{align}
with $\delta=m/P^z$. If the $\tilde f$'s are square integrable and their first order derivatives are continuous, one can show that as $\delta \rightarrow 0$,
\begin{align}
|\tilde f_1(y,P^z)-\tilde f_2(y,P^z)|\sim \delta \ .
\end{align}
Therefore, the ambiguity of different schemes disappears in the large $P^z$ limit.

However, this is still unsatisfactory because it appears that, due to non-perturbative
effects from the linear divergence, the LaMET expansion will be
an expansion in powers of $M/P^z$ instead of $(M/P^z)^2$, which
will significantly reduce the speed of convergence. Here
we consider possible ways to overcome this deficiency. Recall that the LaMET expansion in $(M/P^z)^2$ is
made in the $\MS$ scheme where no linear divergence exists, in a general scheme this expansion might 
contain odd powers in $1/P^z$. Therefore, there is a way to choose $m_0$
such that the condition of the $\overline{\rm  MS}$ scheme
\begin{equation}
      \delta m_{\overline{\rm MS}} = 0
\end{equation}
is met with non-perturbative calculations. 
We shall denote such a value of the subtracted mass
by \begin{equation}
\delta m_c = m_{-1}/a + m_{0c}\ .
\end{equation}
where $m_{0c}$ can be determined by matching the matrix element $\tilde h_1(z, P^z, a)$ on lattice
to the $\overline{\rm MS}$ result when $z$ is small (around $z_S$) and 
QCD perturbative theory works. An alternative strategy
is to vary $m_{0}$ in a certain range near $\Lambda_{\rm QCD}$,
and identify the value $m_{0c}$ for which the linear term in $1/P^z$ in $\tilde f(y, P^z)$ vanishes. 
This is like searching for the critical value of $\kappa_c$ for Wilson
fermions for which a similar power divergent
bare quark mass appears~\cite{Lepage:1992xa}.

The mass-subtracted operator $O(z,a)e^{-\delta m(a)|z|}$
has no power divergence, but still has logarithmic
dependence on $a$.
These remnant logarithmic divergences are independent of $z$ and can be renormalized, in principle, using the auxiliary field method~\cite{Green:2017xeu,Green:2020xco}.
However, a more convenient option in practice is to fix the renormalization constant $Z_{\rm hybrid}(a)$ by directly
matching the renormalized matrix elements of $O(z,a)$ at $z=z_{\rm S}$ from the short and long distance regimes, which is essentially a continuity condition,
\begin{align}
Z_{\rm hybrid} e^{-\delta m |z_{\rm S}|}\langle P| O(z_{\rm S},a) | P \rangle  = \frac{\langle P| O(z_{\rm S},a) | P \rangle }{Z_X(z_{\rm S},a) }\,,
\end{align}
which leads to
\begin{align}
Z_{\rm hybrid}(z_{\rm S},a) = e^{\delta m |z_{\rm S}|}/{Z_X(z_{\rm S},a)  }\,.
\end{align}
In this way, one only has to calculate $\delta m$. Of course, one needs to vary $z_{\rm S}$ to check whether the final result is stable.

The matching coefficient ${\cal C}_{\rm hybrid}$ for the long distance regime is related to that for the X-scheme. For example, if one adopts the $P^z=0$ matrix element for renormalization~\cite{Zhang:2018ggy,Radyushkin:2017lvu,Izubuchi:2018srq}, then
\begin{align}\label{eq:hybridm1}
&{\cal C}_{\rm hybrid}(\alpha, z^2\mu^2, z^2/z_{\rm S}^2) = {\cal C}_{\rm ratio}(\alpha, z^2\mu^2) \nn\\
&\qquad + \delta(1-\alpha){\alpha_sC_F\over 2\pi}{3\over 2} \ln {z^2\over z_{\rm S}^2}\ \theta(|z|-z_{\rm S})\,.
\end{align}
 However, due to the logarithms of $z^2\mu^2$ and $z^2/z_{\rm S}^2$, the above matching coefficient is only valid for $|z| \ll \Lambda_{\rm QCD}^{-1}$, otherwise one has to resum the large logarithms for $|z|\sim \Lambda_{\rm QCD}^{-1}$ by evolving $\alpha_s$ to a highly nonperturabtive regime. Since our ultimate goal is to Fourier transform the final result to obtain the PDF, this will introduce uncontrolled sytematics.

To have a clearer way of separating the perturbative and non-perturbative regimes, we can perform the matching in momentum space, where nothing prevents using the correlations at large $z$, provided that $yP^z$ is sufficiently large. In principle, we should first convert the hybrid scheme to the $\MS$ scheme---where the factorization formula was proven~\cite{Ma:2014jla,Ji:2020ect}---in coordinate space with the conversion factor
\begin{align}\label{eq:hybrid3}
    {\cal Z}(z,z_S,\mu) & = {Z_{\MS}\over Z_{\rm hybrid}} =\frac{\theta(z_{\rm S}\!-\!|z|)}{Z^{\overline{\rm MS}}_{X}(z,\mu)}
    +\frac{\theta(|z|\!-\! z_{\rm S})}{Z^{\overline{\rm MS}}_{X}(z_S,\mu)}\,,
\end{align}
where $1/Z^{\overline{\rm MS}}_{X}$ converts the ``$X$'' scheme to $\MS$, and for \eq{hybridm1}
\begin{align}\label{eq:zmsr}
    Z^{\overline{\rm MS}}_{\rm ratio} = 1 - {\alpha_sC_F\over 2\pi}\left[{3\over 2}\ln{z^2\mu^2\over 4e^{-2\gamma_E}}+ {5\over2}\right]\,.
\end{align}
The conversion factor ${\cal Z}$ is perturbative for all $z$ as it does not include $\ln(z^2)$ at large distance $|z|>z_{\rm S}$. Then we can Fourier transform the $\MS$ quasi-LF correlation and match it to the PDF in momentum space.

Since the scheme conversion is perturbative for all $z$, we can also do the Fourier transform first and directly match the hybrid scheme quasi-PDF to the PDF, and the matching coefficient $C_{\rm hybrid}$ is given by the double Fourier transform from \eq{hybridm1},
\begin{align}\label{eq:hybridm2}
&C_{\rm hybrid}(\xi, \mu^2/(p^z)^2, z_{\rm S}^2\mu^2) = C_{\rm ratio}(\xi, \mu^2/(p^z)^2) \nn\\
&\quad+{\alpha_sC_F\over 2\pi}{3\over 2} \left[-{1\over |1-\xi|_+}+ \frac{2 \text{Si}((1-\xi) \lambda_{\rm S})}{\pi  (1-\xi)} \right]\,,
\end{align}
where $C_{\rm ratio}$ can be found in~\cite{Izubuchi:2018srq}, $\xi=y/x$, and $\lambda_{\rm S}=z_{\rm S}p^z$ with $p^z=xP^z$ being the parton momentum. The plus function is defined as
\begin{align}
	{1\over |1-\xi|_+} &\equiv \lim_{\beta\to0^+}\left[ {\theta(|1-\xi|-\beta)\over |1-\xi|} + 2\delta(1-\xi)\ln\beta\right] \,.
\end{align}

We can also derive the corresponding scheme conversion factor and matching coefficient for using RI/MOM scheme in the short-distance renormalization. In the limit of $-z^2_{\rm S}p^2 \ll 1$, the RI/MOM renormalization factor $Z(z,-p^2)$ for $|z|<z_{\rm S}$ is equal to that of the zero-momentum matrix element, so the results are the same as those in \eqs{hybrid3}{hybridm2}. However, if $-z^2_{\rm S}p^2$ is finite, then one needs to use the results from Refs.~\cite{Constantinou:2017sej,Stewart:2017tvs} to derive the scheme conversion factor and matching coefficient. 
Finally, since the result of the PDF must be independent of the lattice renormalization scheme, we can try different short-distance schemes and check if they are consistent with each other.

In momentum space, the matching coefficient includes the logarithm of $\mu/(yP^z)$ which becomes non-perturbative for $y\sim \Lambda_{\rm QCD}/P^z$. This is consistent with the power counting parameter $\Lambda^2_{\rm QCD}/(yP^z)^2$. Therefore, the nature of the systematic uncertainties is clear, and we can only improve precision at
small $x$ by pushing to higher $P^z$.

\section{Strategy of data analysis at asymptotic distances}
\label{sec:regge}

For finite hadron momentum, lattice calculations of quasi-LF correlations always end up with data at
finite $\lambda_{\rm L}=P^zz_{\rm L}$ where $z_{\rm L}$ is usually smaller than the lattice size due to increasing finite volume corrections and worse signal-to-noise ratios at large $z$. However, to reconstruct the full parton distribution, we need the correlations at all quasi-LF distances.

At finite momentum, the quasi-LF correlation in general has a finite correlation length (in the $\MS$ or hybrid scheme) and exhibits an exponential decay at large $z$. This is similar to the case of density-density~\cite{Burkardt:1994pw} or current-current correlation since the quasi-LF correlation can be viewed as the product of two heavy-to-light currents in the auxiliary field formalism~\cite{Samuel:1978iy,Gervais:1979fv,Arefeva:1980zd,Dorn:1986dt}. As a consequence, its Fourier transform converges fast at finite $z$ or $\lambda$, as compared to that of the LF or twist-2 correlation which only decays algebraically at large $\lambda$ due to the Regge behavior. If $z_{\rm L}$ is large enough such that the quasi-LF correlation falls close to zero, we can do a truncated Fourier transform up to $z_{\rm L}$ to obtain the quasi-PDF, and the resulting systematic uncertainty is negligible compared to other sources.

However, in practical lattice calculations, the choice of $z_L$ is limited by fast-growing errors of quasi-LF correlations. This is particularly true for large hadron momentum.
Thus, when we choose a $z_L$ or $\lambda_{\rm L}$ with a target error, the quasi-LF correlation may still have a sizeable nonzero value at that point. In this case, a truncated Fourier transform will lead to an unphysical oscillation and inaccurate small-$x$ result in the quasi-PDF, which can be formulated as an inversion problem~\cite{Karpie:2019eiq}. Several strategies have been adopted in the literature to address this issue, e.g., the Backus-Gilbert method~\cite{Karpie:2019eiq,Bhat:2020ktg}, neural network and Bayesian reconstructions~\cite{Karpie:2019eiq}, the Gaussian reweighting method that suppresses the long-range correlations~\cite{Ishikawa:2019flg}, the derivative method~\cite{Lin:2017ani} which amounts to doing integration-by-parts and ignoring the boundary terms at the truncation point, or the Bayes-Gauss-Fourier transform which reconstructs a continuous form of the quasi-LF correlation over the whole domain by employing Gaussian process regression~\cite{Alexandrou:2020tqq}. However, the assumptions employed in these strategies are mostly based on mathematical rather than physical reasons. Here we propose to use the knowledge of the asymptotic behavior of quasi-LF correlations and perform a physically motivated extrapolation to $\lambda\!=\!\infty$.
After the extrapolation, one can perform a discrete Fourier transform for $|z|\le z_{\rm L}$, where the discretization error can be studied with the lattice spacing dependence, while for the extrapolated part one can perform the Fourier transform analytically. Therefore, the mathematical inverse problem is solved by physics considerations. 
Although this extrapolation does not provide a first-principle prediction of the small-$x$ PDF, it helps remove the unphysical oscillation and offers a reasonable way to estimate the systematic uncertainties in this region.

Depending on how large the momentum is, we propose to use either the exponential or algebraic decay form for the extrapolation. In the following, we discuss them in detail and describe how to estimate the corresponding uncertainty.

\subsection{Exponential extrapolation at moderately large $P^z$}

\begin{figure}[t]
\centering
\includegraphics[width=\columnwidth]{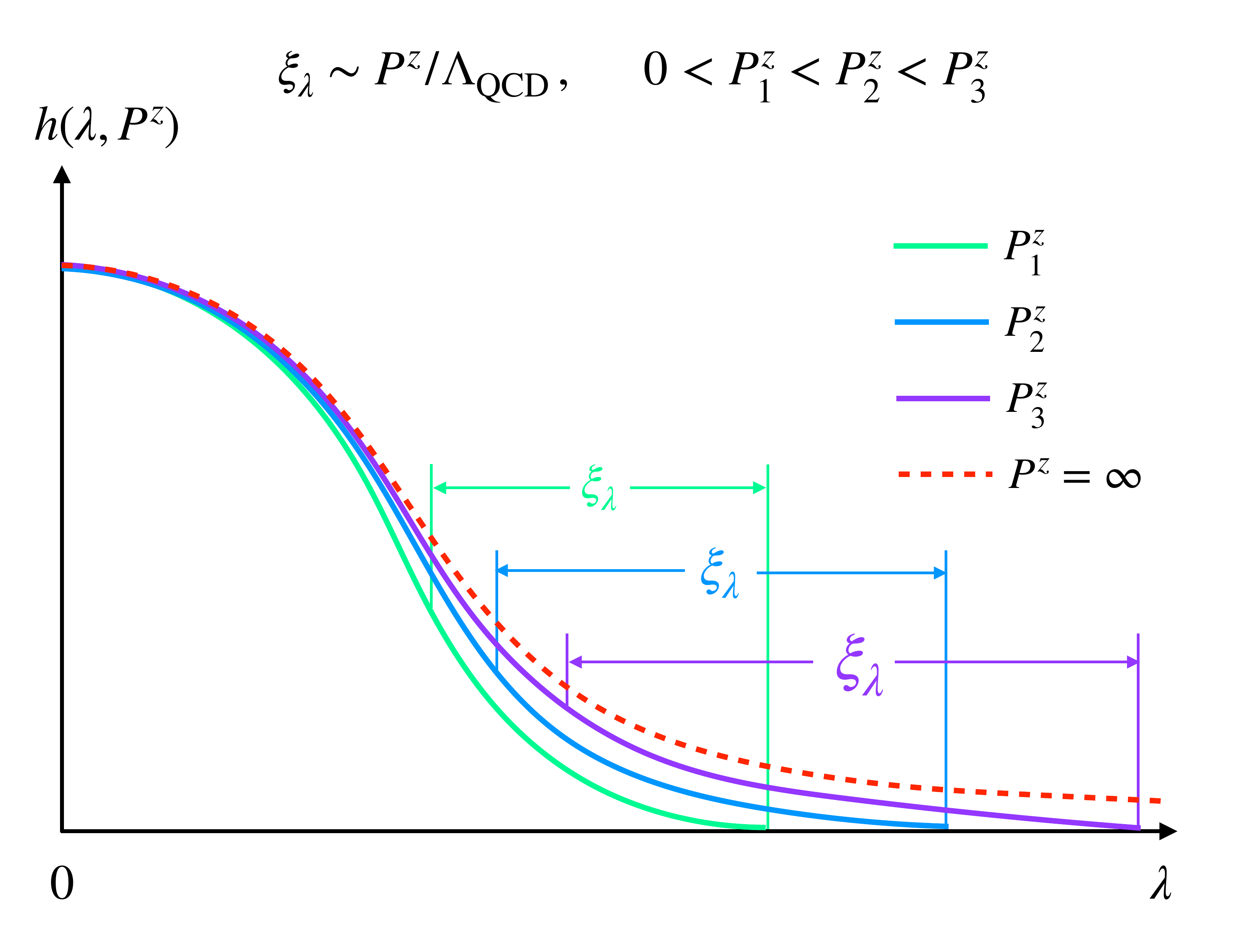}
\caption{Qualitative behavior of the quasi-LF correlation in $\lambda$ space at different $P^z$. For finite $P^z$, at short $\lambda$ the correlation is approximated by the leading-twist contribution and evolves slowly in $P^z$. At large $\lambda$, the correlation starts to exhibit the exponential decay behavior, and both the starting point and correlation length $\xi_\lambda$ increase with respect to $P^z$. In the $P^z\to\infty$ limit, $\xi_\lambda$ approaches infinity and the quasi-LF correlation only includes leading-twist contribution which decays algebraically.}
\label{fig:extr}
\end{figure}

As mentioned above, for a moderately large momentum $P^z$, the quasi-LF correlations in general have a finite correlation
length $\xi_z \sim 1/\Lambda_{\rm QCD}$ in the coordinate $z$ space or
$\xi_\lambda \sim P^z/\Lambda_{\rm QCD}$ in the LF distance $\lambda$ space. This
is due to the confinement property of non-perturbative QCD and spacelike nature of the correlation. The finite correlation length is associated with an exponential decay $\exp(-z/\xi_z)$, which becomes significant at large $z$. Before the quasi-LF correlation exhibits the exponential decay behavior, it is dominated by the leading-twist contribution which evolves slowly in $P^z$. In the $\lambda$ space, the quasi-LF correlations at different $P^z$ can be qualitatively described by \fig{extr}. As $P^z$ increases, the quasi-LF correlation evolves closer to leading-twist contribution, and starts to exhibit the exponential decay at larger $\lambda$ values. In the limit of $P^z\to\infty$, $\xi_\lambda$ approaches infinity and the quasi-LF correlation only includes the leading-twist contribution that decays algebraically at large $\lambda$.

Therefore, when $P^z$ is not very large, e.g., about 2--5 GeV for the proton, we propose to use the exponential decay form $\sim e^{-\lambda/\xi_\lambda}$ to do the extrapolation (although some algebraic behavior can be added on the top to better represent the $\lambda$-dependence of the quasi-LF correlations). Note that to make the extrapolation under control, it is critical for the lattice data to exhibit the exponential decay before the error becomes too big. This shall be achieved with larger lattice volume and/or higher statistics of measurements, which are feasible for contemporary computing resources. In the ideal case, at large enough $z$ or $\lambda$, the quasi-LF correlation would become practically zero within the target error, and the extrapolation will barely affect the final result except for the extremely small-$x$ region. In more practical scenarios, the lattice data shows the exponential drop, but still has a statistically significant nonzero value at $\lambda_{\rm L}$, then we can perform the exponential extrapolation to $\lambda=\infty$.

Note that although the exponential form is physically motivated, it remains unclear at what values of $z$ or $\lambda$ the lattice data should be included for the fitting, as the large-$z$ data still includes leading-twist contribution which may obscure the result. Namely, one may fit to different values of the correlation length $\xi_\lambda$ with different choices of the fitting range. Nevertheless, the variation in $\xi_\lambda$ will mainly affect the region with very small $x$, which are anyway less predictive due to power corrections. Therefore, it is not a prerequisite to fit $\xi_\lambda$ precisely. Instead, one should utilize this property by varying the fitting range, e.g., within $z_{\rm L}-5a\le z \le z_{\rm L}$, and test the stability of the final result with different $\xi_{\lambda}$.

Last but not the least, the Fourier transform of an exponentially decaying correlation always leads to a finite quasi-PDF at $x=0$, which is different from the Regge behavior of PDFs at small $x$. Besides, since the PDF at large $x$ ($x\to1$) is also sensitive to the long-range LF correlation which decays algebraically, the quasi-PDF shall deviate from the PDF in this region, too. These indicate the significance of power corrections in the end-point regions as $x\to0$ and $x\to1$, which gives us the hint on how to estimate the systematic uncertainty from the exponential extrapolation. To be specific, we can perform an algebraic extrapolation (see the section below) for the same range of data, which is essentially equivalent to ignoring all the power corrections at large $z$, and choose its difference to the exponential extrapolation as the error. One can anticipate that this estimate will lead to increasingly large systematic errors as $x$ approaches the end points, which is consistent with the accuracy of the momentum-space expansion.

\subsection{Algebraic extrapolation at very large $P^z$}

When $P^z$ becomes very large with future lattice resources,
 $\xi_\lambda$ also becomes very large and it will take larger $\lambda$ values to see the exponential decay in lattice data. We expect that the decay behavior follows more like an algebraic law rather than an exponential one as $\lambda\sim\lambda_{\rm L}$. In other words, the quasi-LF correlation is very close to the leading-twist correlation since the power corrections for $\lambda \le \lambda_{\rm L}$ are expected to be well suppressed. In this case, we can use an algebraic form to extrapolate to $\lambda=\infty$.

The algebraic decay of leading-twist correlation is a consequence of its infinite correlation length $\xi_\lambda$, and is associated with the asymptotic Regge behavior~\cite{Regge:1959mz}. At small $x$, it is well-known that parton distributions behave asymptotically like $x^a$, as suggested by Regge theory. For the non-singlet combination, the leading Regge trajectory indicates that $a\sim-1/2$. For the singlet combination, its mixing with gluon distributions under evolution makes things more subtle. In the so-called soft pomeron model, one has $a\sim -1$. However, scattering data at large momentum transfer indicate a more singular asymptotic behavior, reflecting the potential need for a contribution of the hard pomeron~\cite{Devenish:2004pb}. At large $x$, the asymptotic behavior is dictated by the quark counting rules~\cite{Brodsky:1973kr}. As $x\to 1$, the hadron momentum is carried by the struck quark and no momentum is left for other spectator partons. The asymptotic behavior is then predicted to be $(1-x)^b$, where $b=2n_s-1+2|\Delta S^z|$ with $n_s$ being the minimum number of spectator partons and $\Delta S^z$ the difference of the spin projections for the struck parton and the parent hadron~\cite{Devenish:2004pb,Brodsky:2005wx}. For example, for a valence quark in the proton $b=3\, (5)$ if the struck quark has helicity parallel (antiparallel) to the proton as $n_s=2$ and $|\Delta S^z|= 0\, (1)$, while for the pion one has $b=2$ since $n_s=1$ and $|\Delta S^z|=1/2$. The above features have been widely used in global fits of PDFs, where one parameterizes the PDFs such that they behave as $x^a$ for $x\to 0$ and $(1-x)^b$ for $x\to 1$ and fit the powers $a, b$ to a large variety of experimental data.
The role of such a power law behavior in global fits has been examined in detail in Refs.~\cite{Ball:2016spl,Nocera:2014uea}.

When Fourier transformed to coordinate space, the asymptotic behavior described above implies that the correlation in the longitudinal space decays algebraically as $\lambda^{-\alpha}$ ($\alpha$ is a positive number related to $a, b$) rather than exponentially, and thus has an infinite correlation length. A similar algebraic decay behavior was also observed in a recent analysis of the LF wave functions~\cite{Brodsky:1997de} when Fourier transformed to conjugating coordinate space~\cite{Miller:2019ysh}.

To see how the asymptotic behavior can help with the extrapolation of quasi-LF correlations at large momentum, let us begin with the following simple form of PDFs that incorporates the $x\sim 0, 1$ behavior,
\begin{equation}
     x^a (1-x)^b \ .
\end{equation}
The coordinate space matrix element can be defined as
\begin{equation}
h(\lambda)=\int_0^1 dx \, e^{i x\lambda}x^a (1-x)^b,
\end{equation}
from which it follows that at large $\lambda$
\begin{align}\label{asympt_coord}
h(\lambda)\sim\frac{\Gamma(1+a)}{(-i|\lambda|)^{a+1}}+e^{i\lambda}\frac{\Gamma(1+b)}{(i|\lambda|)^{b+1}} \ ,
\end{align}
whose real (imaginary) part is even (odd) in $\lambda$, ensuring that parton distributions are real functions in momentum space. Therefore, the conjugate LF correlations behave at large $\lambda$ as $\lambda^{-\alpha(a,b)}$ with
\begin{align}
\alpha(a,b)=\text{\rm min}(a+1,b+1).
\end{align}
In most cases we are interested in, $\alpha(a,b)=a+1$. Applying the matching in Eq.~(\ref{matching}) converts the light-cone correlations to quasi-LF correlations, and also induces logarithmic corrections to the asymptotic behavior. In regions where the factorization is valid, such corrections can be resummed as $\sim \exp(\gamma \ln z^2\mu^2)=(\lambda^2\mu^2/(p^z)^2)^\gamma$ to leading logarithmic (LL) accuracy, which modifies the asymptotic behavior of the quasi-LF correlation as
\begin{align}\label{asymphtilde}
\tilde h(\lambda,z)\sim e^{\gamma \ln z^2\mu^2}\frac{1}{|\lambda|^{\alpha(a,b)}}\sim |\lambda|^{-\alpha(a,b)+2\gamma }  \,.
\end{align}
This provides a useful approximation to the quasi-LF correlations at large $\lambda$ with sufficiently large $P^z$, and is consistent with previous discussions based on the correlation length.
Now we can use the following algebraic form to extrapolate the quasi-LF correlation to infinite $\lambda$ (taking $\lambda>0$ as an example)
\begin{equation}\label{extrap}
\frac{c_1}{(-i \lambda)^{d_1}}+e^{i\lambda}\frac{c_2}{(i \lambda)^{d_2}},
\end{equation}
which accommodates the two different structures in Eq.~(\ref{asympt_coord}). The parameters $c_i, d_i$ can be fitted in the same way as that in the exponential extrapolation. Finally, we can use the uncertainty in these parameters to estimate the systematic error from extrapolation.

By supplementing lattice data with the above extrapolation strategy, we expect the final PDF result to be free of unphysical oscillation and converge better to the physical region $0<x<1$.

\section{Large Momentum Vs. Short Distance Expansion}

The Euclidean correlator in Eq. (5) introduced in Ref. \cite{Ji:2013dva}
has also been considered in coordinate-space factorizaton (CSF)~\cite{Radyushkin:2017cyf}, which was
introduced in an early work on meson DAs with current-current correlators
\cite{Braun:2007wv} (see also~\cite{Ma:2017pxb}).
The correlator can be factorized in terms of the LF correlations
with expansion parameter $(z\Lambda_{\rm QCD})^2$. The formalism is
naturally suited for calculating moments of PDFs or short-distance
LF correlations. To obtain the full parton physics, however, one has to
simultaneously consider the constraint on the external momentum $P^z\sim 1/z \gg \Lambda_{\rm QCD}$.
This is identical to the observation in Ref. \cite{Ji:2014gla}: One must use
large momenta to capture the full dynamical range of PDFs, which requires information on
long-range correlations in $\lambda$.
Despite their formal equivalence~\cite{Ji:2017rah,Izubuchi:2018srq,Radyushkin:2017lvu},
some analytical matching calculations might
more conveniently be done in coordinate space. Not surprisingly, the same LaMET
lattice data are needed for a CSF analysis to get the PDFs.
Nominally, CSF can also admit data at small $P^z$, but the same information is
contained already in large $P^z$ data at smaller $z$.

The CSF expansion is formulated in terms of
the Euclidean distance $z$, which is required to be small, i.e.,
\begin{equation}
         z\ll 1/\Lambda_{\rm QCD}\ ,
\end{equation}
to ensure the validity of perturbation theory and leading-twist dominance.
Assuming the largest $z$ for the leading-twist
approximation to be $z_{\rm LT}$ (say, the value of $z$ for which the higher-twist contribution
is at the level $\sim20\%$), then the small
expansion parameter is $(z_{\rm LT}\Lambda_{\rm QCD})^2\ll 1 $ when
potential linear divergences are subtracted
before the expansion is made. Therefore, only the matrix element
of $O(z)$ within the range $[0, z_{\rm LT}]$ has a simple interpretation
in terms of leading-twist parton physics.

An interesting question is then: What is the value of $z_{\rm LT}$?
If we take $\Lambda_{\rm QCD}\sim 300$ MeV, and  $z_{\rm LT}\Lambda_{\rm QCD}=1/2\sim 1/3$ as
a small parameter, then the estimate is that $z_{\rm LT}$ is around $0.25 \sim 0.33$ fm. An
upper limit is probably 0.4 fm. A good estimate of $z_{\rm LT}$ can be provided by comparing the matrix element $\langle P=0| O(z)|P=0\rangle$ or $\langle \Omega| O(z)|\Omega\rangle$, both of which have been proposed to renormalize the bare quasi-LF correlation~\cite{Radyushkin:2017cyf,Braun:2018brg,Li:2020xml}, to the leading-twist contributions in their OPE.
	
Let us take the zero-momentum matrix element  for the isovector case as an example. In the $\MS$ scheme, it has a short distance expansion of the form~\cite{Radyushkin:2017lvu,Izubuchi:2018srq}
\begin{align}
			\tilde{h}(z,\mu,P^z\!=\!0)&= {1\over 2M}\langle P=0| \bar{\psi}(z)\gamma^0 W(z,0)\psi(0)|P=0\rangle \nn\\
			& = c_0(\mu^2 z^2) a_0 + {\cal O}(z^2\Lambda_{\rm QCD}^2)\,,
\end{align}
where $W(z,0)$ is a spacelike straight gauge link. Here $\mu$ is the $\MS$ renormalization scale, and $a_0=1$ is the conserved lowest moment of the correpsonding twist-2 PDF. The one-loop Wilson coefficient $c_0=1/  Z^{\overline{\rm MS}}_{\rm ratio}$ with $Z^{\overline{\rm MS}}_{\rm ratio}$ given in \eq{zmsr}~\cite{Izubuchi:2018srq},
and the two-loop result can be found in Ref.~\cite{Li:2020xml}.

According to \eq{ren}, the mass-subtracted matrix element includes logarithmic divergences that are independent of $z$ and should not constitute significant corrections in lattice perturbation theory. Therefore, we can roughly approximate its OPE by replacing $\mu$ with $1/a$,
\begin{align}\label{eq:subtracted}
	e^{-\delta m |z|}\tilde{f}(z,a,P^z\!=\!0) \!=\! c_0(z^2/a^2) \!+\! {\cal O}(z^2\Lambda_{\rm QCD}^2, a^2\!/\!z^2)\,,
\end{align}
where the lattice discretization effects are expected to be of ${\cal O}(a^2/z^2)$.
Since the lattice matrix elements are convergent as $z\to0$, which is contrary to the logarithmically divergent behavior in the $\MS$ OPE, we expect the above approximation to be reliable within the range $a < |z| < z_{\rm LT}$ where the discretization and higher-twist effects are both suppressed. Note that lattice OPE is usually complicated by the broken Lorentz symmetry and operator mixings. Nevertheless, since for $P^z=0$ the only leading-twist contribution comes from the conserved vector current, we can ignore such effects here.

\begin{figure}[t]
\centering
\includegraphics[width=1\columnwidth]{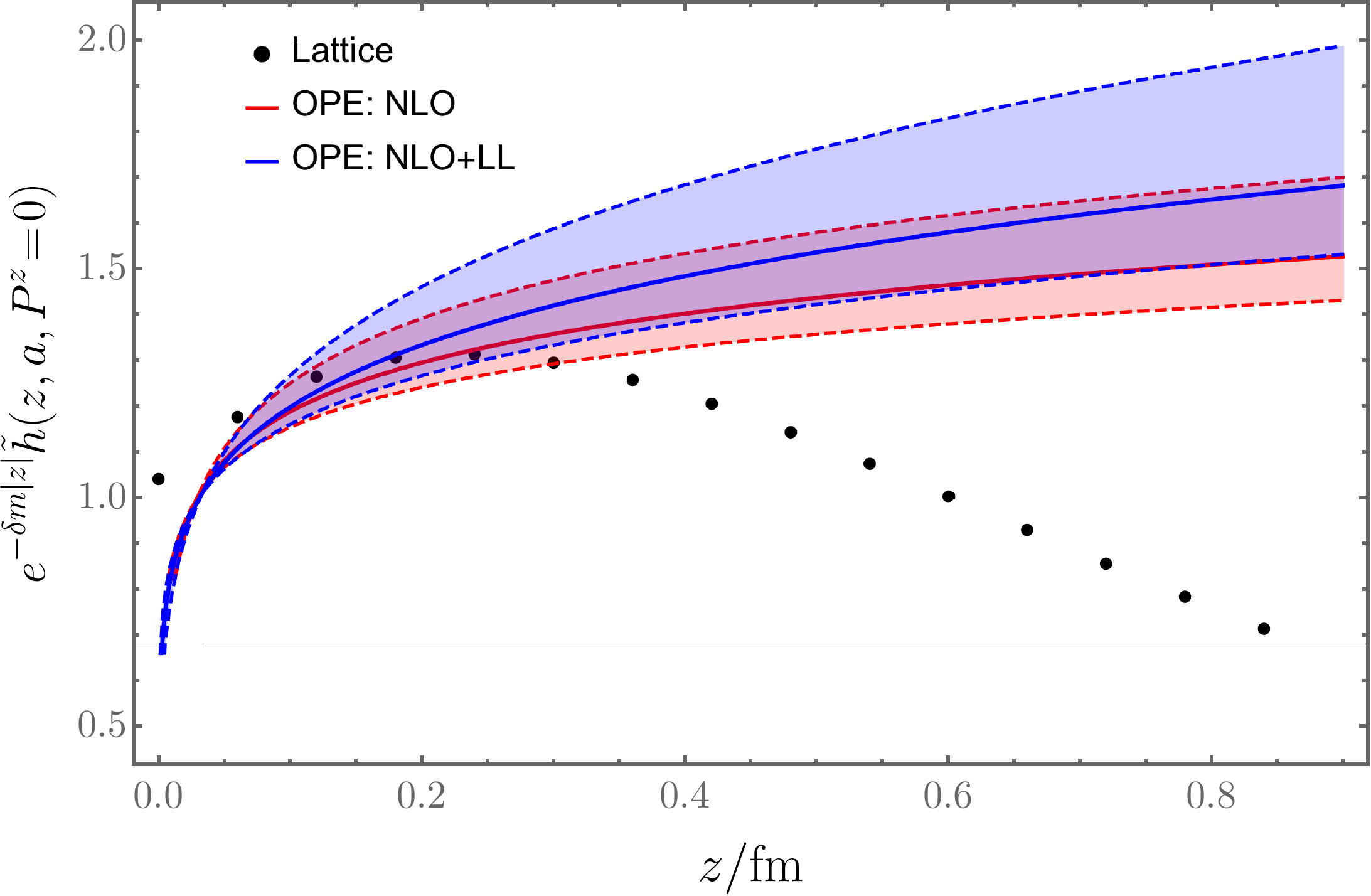}
\caption{Comparison of the mass-renormalized pion lattice matrix element~\cite{Gao:2020ito} and its leading-twist approximation with NLO and NLO$+$LL corrections for fixed $\alpha_s$, respectively.
The strong coupling is $\alpha_s(1/a)=0.242$, and the error band was obtained by varying $\alpha_s$ from $\alpha_s(1/(2a)$ to $\alpha_s(2/a)$. This shows that the leading-twist approximation becomes unreliable at $z_{\rm LT}\sim 0.3$ fm.}
\label{fig:ope}
\end{figure}

In \fig{ope} we plot the mass-renormalized pion lattice matrix element. The bare lattice matrix element comes from a recent calculation of the pion valence PDF on an ensemble with $a=0.06$ fm and pion mass $m_\pi=300$ MeV~\cite{Gao:2020ito}. On the same lattice ensemble, the Wilson-line mass correction $\delta m$ was fitted from the quark-antiquark potential~\cite{Izubuchi:2019lyk}, and its value is given in lattice units as $a \delta m = -0.1568$. The leading-twist contribution is plotted with next-to-leading-order (NLO) corrections and NLO correction plus LL resummation for fixed $\alpha_s$,
\begin{align}
	\left[ 1 + {\alpha_s(1/a)C_F\over 2\pi}  {5\over2}\right]\left({z^2\over 4e^{-2\gamma_E}a^2}\right)^{{\alpha_s(1/a)C_F\over 2\pi}{3\over 2}}\,.
\end{align}
Here we choose $\alpha_s$ in the $\MS$ scheme at scale $1/a$ as the input for OPE, which should allow for better convergence than the bare lattice coupling~\cite{Lepage:1992xa}. To estimate the uncertainty from the choice of $\alpha_s$, we vary the $\MS$ scale from $1/(2a)$ to $2/a$. Though a standard procedure of improvement shall be performed to define $\alpha_s$ on the lattice, we expect that it will not alter the following conclusion.

As one can see, for $z\le a$, the lattice result is significantly different from the leading-twist approximations due to discretization effects. As $z$ increases, the agreement becomes better. However, for $z\ge 0.3$ fm, the lattice result starts to deviate dramatically from the leading-twist approximations, showing that the higher-twist contributions become significant. Therefore, we can roughly estimate that $z_{\rm LT}\sim 0.3$ fm.
	
One can also look at the case of the better established heavy-quark potential. It is well-known that the static heavy-quark potential receives both perturbative and non-perturbative contributions. The  perturbative static potential is known up to ${\rm N^3LO}$ level~\cite{Smirnov:2009fh}
and can be expressed in terms of the QCD running coupling constant. In Refs.~\cite{Bazavov:2012ka,Bazavov:2014soa,Bazavov:2019qoo}, the running coupling constant has been extracted from lattice calculation of the static energy at short distances. The ${\rm N^3LL}$ perturbative result agrees well with lattice data up to $r \sim 0.2$ ${\rm fm}$. However, it is well-known that the perturbative series for the static potential suffers from a renormalon ambiguity~\cite{Beneke:1998rk,Hoang:1998nz} and breaks down at large distance. The non-perturbative heavy-quark potential can be simulated using lattice QCD, and is well-known to be dominated by the linear term of the form $\sigma r$ at large distance. Phenomenologically, the static potential can be well approximated by the linear+Coulumb QCD static potential, $V(r)=-e/r+\sigma r$ where $e= 0.25\sim 0.5$ while $\sqrt{\sigma} \approx 477$ ${\rm MeV}$~\cite{Eichten:1974af,Bali:2000gf,Aubin:2004wf}. When the perturbation theory is about to break down, the perturbative contribution $-e/r$ and the confining contribution $\sigma r$ should be of the same order of magnitude, which determines $r_c \approx\sqrt{e/\sigma}$ to be around $0.2\sim 0.3$ ${\rm fm}$. This is consistent with the result in Refs.~\cite{Bali:1999ai,Necco:2001gh,Bazavov:2012ka}. The boundary $z_{\rm LT}$ should be of the same order of magnitude.

\begin{figure}[t]
\includegraphics[width=1\columnwidth]{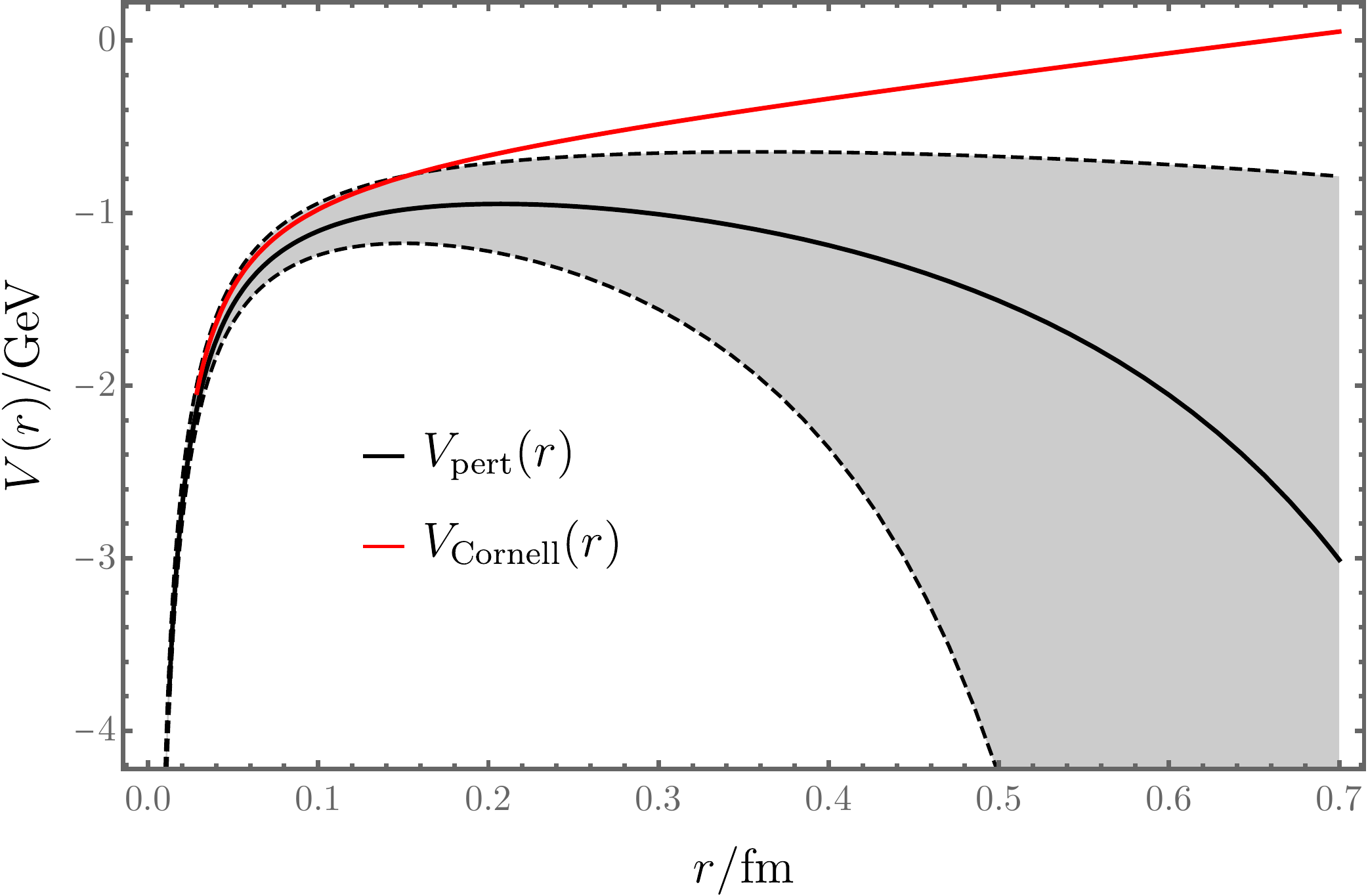}
\caption{A comparison between the perturbative potential in $\alpha_s^3$ order to the linear+Coulumb QCD static potential taken from~\cite{Bali:2000gf,Aubin:2004wf}. The shaded area corresponds to the perturbative prediction with the renormalization scale $\mu$ ranging from $r/2$ to $3r/2$. The black line corresponds to $\mu=1/r$. From the figure, it is clear that beyond $0.2$ to $0.3$ ${\rm fm}$, the perturbative potential starts to deviate from the full non-perturbative results and becomes unreliable.}
\label{fig:poten}
\end{figure}

With the estimated $z_{\rm LT}$ above, one can also define
\begin{equation}
       \lambda_{\rm LT} = P^zz_{\rm LT},
\end{equation}
then the matrix element for the quasi-LF distance $[0, \lambda_{\rm LT}]$
can be used to extract parton distributions with the matching formula in \eq{coordfact}~\cite{Radyushkin:2017cyf,Ji:2017rah,Izubuchi:2018srq}.
Thus, the coordinate-space approach is useful for extracting the LF correlation functions in a limited range, with the LF distance ranging between $[0,\lambda_{\rm LT}]$.

The CSF approach has also been used for products of currents made of
quark bilinears~\cite{Detmold:2005gg,Braun:2007wv,Bali:2017gfr,Bali:2018spj,Ma:2017pxb,Detmold:2018kwu,Sufian:2019bol,Sufian:2020vzb}.
Renormalization of power divergences in the
CSF expansion for the quark and gluon blinears with Wilson line
is easier to handle. In particular, a version of the ratio method
which  divides by the matrix element at zero momentum, can be used to
eliminate the power divergences in the lattice matrix elements~\cite{Radyushkin:2017cyf,Orginos:2017kos}.
On the other hand, the current products can also be used
in LaMET expansion after Fourier transforming into momentum
space~\cite{Ji:2020ect}.

However, the CSF does not allow for directly calculating the $x$-distribution, because one needs
the LF correlation at all LF distances. The requirement for $z\le z_{\rm LT}$ makes it unfeasible
to reach large $\lambda$ values for the Fourier transform with the largest momentum on contemporary lattice resources. Thus, to reconstruct
the PDF, one has to parameterize the functional form of the $x$ dependence, just like that
in the phenomenological fits, and then
inverse Fourier transform it to the coordinate space and fit to a limited range
of LF correlations~\cite{Orginos:2017kos,Karpie:2018zaz,Joo:2019bzr,Joo:2019jct,Sufian:2019bol,Joo:2020spy,Sufian:2020vzb,Bhat:2020ktg,Gao:2020ito,Fan:2020cpa}.
This process is hardly under control, because it is difficult to estimate
the systematic uncertainty from the parameterization or assumptions of the PDF. As evident in the fits performed in literature so far~\cite{Joo:2019bzr,Joo:2019jct,Sufian:2019bol,Joo:2020spy,Gao:2020ito,Fan:2020cpa}, either the errors in the end-point regions become smaller and smaller, or the errors shrink to almost zero for certain moderate values of $x$. These imply unaccounted systematics from the artifacts of the particular model used, which is also reflected by their inconsistency with global fits that use similar parameterizations.

From a different angle, the above practice amounts
to postulating (or modeling) certain correlation between short- and long-
distance behaviors of the LF correlations. Such a postulation
has no first-principles foundation, and it can happen that the lattice data in
the limited range of $\lambda$ be fitted equally well by more than one parameterizations which
have completely different asymptotic behaviors~\cite{Sufian:2020vzb}.

Despite the difficulty in providing a controlled calculation of the $x$-distribution, the CSF method allows for model-independent extraction of the Mellin moments using the OPE. Nevertheless, with limited range of $\lambda$, the LF correlations will be sensitive to only the lowest ones, which is related, but not in a direct one-to-one correspondence, to the predictive power in the $x$-space.

To make a more direct comparison between the momentum space and coordinate space approaches, let us consider the following example. Assume that the quasi-LF correlations defining the quasi-PDFs behave like
\begin{align}
\tilde h(\lambda,z)=h(\lambda)e^{-m|z|} \ ,
\end{align}
with $h(\lambda)$ being the light-cone correlator. The exponential $e^{-m|z|}$ with $m\sim \Lambda_{\rm QCD}$ is used to model higher-twist contributions.
From this equation, it is clear that if one stays in position space, the CSF is accurate only when $z\ll 1/m$. The available range of $\lambda$ is therefore much smaller than $P^z/m$, which indicates that the number of moments one can access is much less than $P^z/m$.

From the discussion above, it is clear that the momentum and coordinate space expansions are different expansion schemes. Even though they are equivalent in the infinite momentum limit, they are different at finite
momentum $P^z$.
In the latter, the information is filtered directly
in coordinate space. One gets parton correlations
in a finite range of LF distance which correspond to
the number of moments controlled by $1/P^z$. In contrast, the former uses all the coordinate space information, filtering
higher-twist physics in momentum space through $1/(y^2(P^z)^2)\ll 1$ and $1/((1-y)^2(P^z)^2)\ll 1$.
Therefore, one gets partron distributions in an interval of $x$
with systematic control of errors, which can be directly compared
with experimental data.

Finally, we remark that the relative size of the perturbative correction to the quasi-PDF depends on $x$, where one usually observes larger corrections in the small- and large-$x$ regions. On the other hand, although the size of the perturbative corrections in the coordinate space is usually small for finite $\lambda$, it can still lead to significant corrections in the end-point regions in momentum space.

\section{Conclusion}
\label{sec:conclusion}

To conclude, we have discussed some further subtleties in renormalization
and matching of the quasi-LF correlations on lattice. We proposed a hybrid renormalization procedure to treat the short and long distance correlations separately. The short distance correlations can be renormalized by dividing the same correlator sandwiched in different external states, whereas the long distance ones are renormalized using the Wilson line mass renormalization with a continuity condition to match the short distance region. In this way, we avoid introducing extra non-perturbative effects at large distance in the renormalization stage. We also proposed how to extrapolate to large quasi-LF distance beyond the reach of lattice simulations by utilizing the asymptotic long-range behavior of the correlations, thus avoiding truncations in the ensuing Fourier transform. We finally compared the large-momentum expansion with the CSF approach when applied to LaMET data, showing that the former is a systematic expansion to extract the $x$-dependence of PDFs, whereas the latter is not.
Our proposal here has the potential to greatly improve current computational strategies in lattice applications of LaMET.

\section*{Acknowledgments}

We thank the European Twisted Mass Collaboration and the Brookhaven/Stony Brook University Lattice Group for providing the lattice matrix elements of nucleon and pion. We also thank MILC collaboration for providing the configurations, and J. Hua and Y. Huo, A. Kronfeld, C. Monahan, O. Philipsen, and A. Pineda for valuable discussions and communications. XJ is partially supported by the U.S. Department of Energy under Contract No. DE-SC0020682
and Center for Nuclear Femtography, Southeastern Universities Research Associations in Washington DC.
YZ is supported by the U.S. Department of Energy under award number DE-SC0012704, and within the framework of the TMD Topical Collaboration. JHZ is supported in part by National Natural Science Foundation of China under Grant No. 11975051, and by the Fundamental Research Funds for the Central Universities. AS is supported by SFB/TRR-55. WW is supported in part by Natural Science Foundation of China under grant No. 11735010, 11911530088, by Natural Science Foundation of Shanghai under grant No. 15DZ2272100. YBY is supported by the Strategic Priority Research Program of Chinese Academy of Sciences, Grant No. XDC01040100.


\end{document}